\documentclass[conference]{IEEEtran}
\IEEEoverridecommandlockouts
\usepackage{cite}
\usepackage{amsmath,amssymb,amsfonts}
\usepackage{algorithmic}
\usepackage{graphicx}
\usepackage{textcomp}
\usepackage{xcolor}
\usepackage{url}
\usepackage{array}
\usepackage{mathtools}
\def\BibTeX{{\rm B\kern-.05em{\sc i\kern-.025em b}\kern-.08em
    T\kern-.1667em\lower.7ex\hbox{E}\kern-.125emX}}
\begin{document}


\title{Empowering 9-1-1 Calltaking Training with Generative AI: Experiences and Lessons Learned}


\author{
\IEEEauthorblockN{Zirong Chen}
\IEEEauthorblockA{\textit{College of Connected Computing} \\
\textit{Vanderbilt University}\\
Nashville, TN \\
zirong.chen@vanderbilt.edu}
\and
\IEEEauthorblockN{Meiyi Ma}
\IEEEauthorblockA{\textit{College of Connected Computing} \\
\textit{Vanderbilt University}\\
Nashville, TN \\
meiyi.ma@vanderbilt.edu}
}

\maketitle

\begin{abstract}
Emergency call-takers form the first operational link in public safety response, handling over 240 million calls annually while facing a sustained training crisis: staffing shortages exceed 25\% in many centers, and preparing a single new hire can require up to 720 hours of one-on-one instruction that removes experienced personnel from active duty. Traditional training approaches struggle to scale under these constraints, limiting both coverage and feedback timeliness. In partnership with Metro Nashville Department of Emergency Communications (MNDEC), we designed, developed, and deployed a GenAI-powered call-taking training system under real-world constraints\footnote{This work is a collaboration between the Vanderbilt Research Team and the Metro Nashville Government, including Metro Nashville Information Technology Services and the Metro Nashville Department of Emergency Communications. 
We sincerely thank our partners for their exceptional efforts in providing domain expertise and professional evaluation to support a responsible design, development, and deployment process. We are especially grateful to the IT support, training, quality assurance, and management teams at the Metro Nashville Department of Emergency Communications.} 
Over six months, deployment scaled from initial pilot to 190 operational users across 1,120 training sessions, exposing systematic challenges around system delivery, rigor, resilience, and human factors that remain largely invisible in controlled or purely simulated evaluations. By analyzing deployment logs capturing 98,429 user interactions, organizational processes, and stakeholder engagement patterns, we distill four key lessons, each coupled with concrete design and governance practices. These lessons provide grounded guidance for researchers and practitioners seeking to deliver AI-driven training systems in safety-critical public sector environments where practical constraints fundamentally shape human-centric design.
\end{abstract}



\section{Introduction}
\label{sec:intro}

Emergency call-takers handle over 240 million calls annually in the United States \cite{apco-nena-ans-1-107-1-2015}, serving as critical first responders who coordinate multi-agency emergency response under extreme time pressure \cite{chen2024auto311}. These operators make life-or-death decisions navigating complex protocol trees while managing distressed callers, yet face an escalating training crisis. Centers nationwide report staffing shortages exceeding 25 percent, with some jurisdictions approaching 40 percent deficits \cite{iaed2023}. Training each operator requires up to 720 hours of intensive instruction, removing experienced personnel from active duty for months \cite{apco-ans-3-103-2-2015}. Manual quality assurance covers fewer than 10 percent of calls, creating significant delays between performance and feedback \cite{iaed2023, riikonen2023, chen2026pace}. 
GenAI systems promise scalable solutions through automated scenario generation, realistic caller simulation, and consistent performance assessment \cite{beham2023artificial, hong2025exploring}. However, deploying GenAI within government emergency centers exposes challenges absent from controlled research settings. Prior work identifies adoption barriers in public sector AI including infrastructure limitations, organizational resistance, and skills gaps \cite{campion2020challenges}. Responsible AI frameworks offer governance principles emphasizing transparency, fairness, and accountability \cite{jobin2019global}. Yet existing literature lacks longitudinal empirical evidence of sustained deployment navigating the embedded constraints characteristic of safety-critical government operations. The path from technical prototype to operational routine requires addressing infrastructure, psychological, and governance \textbf{\textit{challenges}} that emerge at production scale: how to deliver systems when bridging knowledge gaps between AI teams and domain experts, how to ensure rigor under safety-critical constraints, how to build resilience through feedback-driven evolution, and how to maintain engagement through calibrated difficulty.



We address this gap through a longitudinal deployment of a GenAI-powered training system with Metro Nashville Department of Emergency Communications (MNDEC). The system automates two training functions: generating realistic emergency scenarios across 57 incident types with diverse caller profiles, and assessing trainee performance against 1,651 protocol requirements translated from natural language to formal specifications. It operates through telephony interfaces where foundation models role-play callers, interact with trainees via audio, and deliver just-in-time debriefing upon session completion. After six months, deployment scaled from an initial pilot across 18 instrumented workstations to 190 operational users completing 1,120 training sessions. Our analysis draws from deployment logs capturing 98,429 user interaction events and 11,129 system events, 5,244 minutes of audio recordings, and development artifacts documenting iterative co-design with stakeholders.





\textbf{Contributions:} This paper shares lessons learned from integrating AI in high-stakes governmental training under sustained operational use. Our contributions are empirically-derived insights from longitudinal production deployment that enable responsible AI adoption in safety-critical public sector environments. We focus on socio-technical dynamics rather than technical system architecture, recognizing that organizational factors, psychological responses, and embedded constraints often determine adoption success more than algorithmic performance alone. Our guidance applies across governmental AI deployments where practical constraints, psychological factors, and organizational dynamics shape adoption success. Specifically, we distill four lessons with actionable practices: (1) \textit{Bridging knowledge gaps} through iterative delivery that transforms domain experts from auditors into co-architects, (2) \textit{Enhancing system rigor} by combining formal verification methods with modularized LLM capabilities, (3) \textit{Building system resilience} through triangulated feedback that distinguishes genuine failures from stress-induced misattributions, and (4) \textit{Designing for constructive challenge} that maintains necessary difficulty while supporting engagement through productive struggle.

\noindent \textbf{\textit{Practitioner Collaboration.}} This work is supported by partnership with MNDEC. Domain experts from training, quality assurance, and operations contributed throughout: validating motivating observations under national and local contexts, co-specifying system improvements and refinements, reviewing system logs, and participating in system evaluation.

\section{Operational Context}
\label{sec:background}

This section describes the deployed system's operational workflow (\ref{sec:workflow}) and the iterative delivery approach (\ref{sec:ddd}). 

\subsection{System Overview}
\label{sec:workflow}


\begin{figure}[t]
  \centering
  \includegraphics[width=\columnwidth]{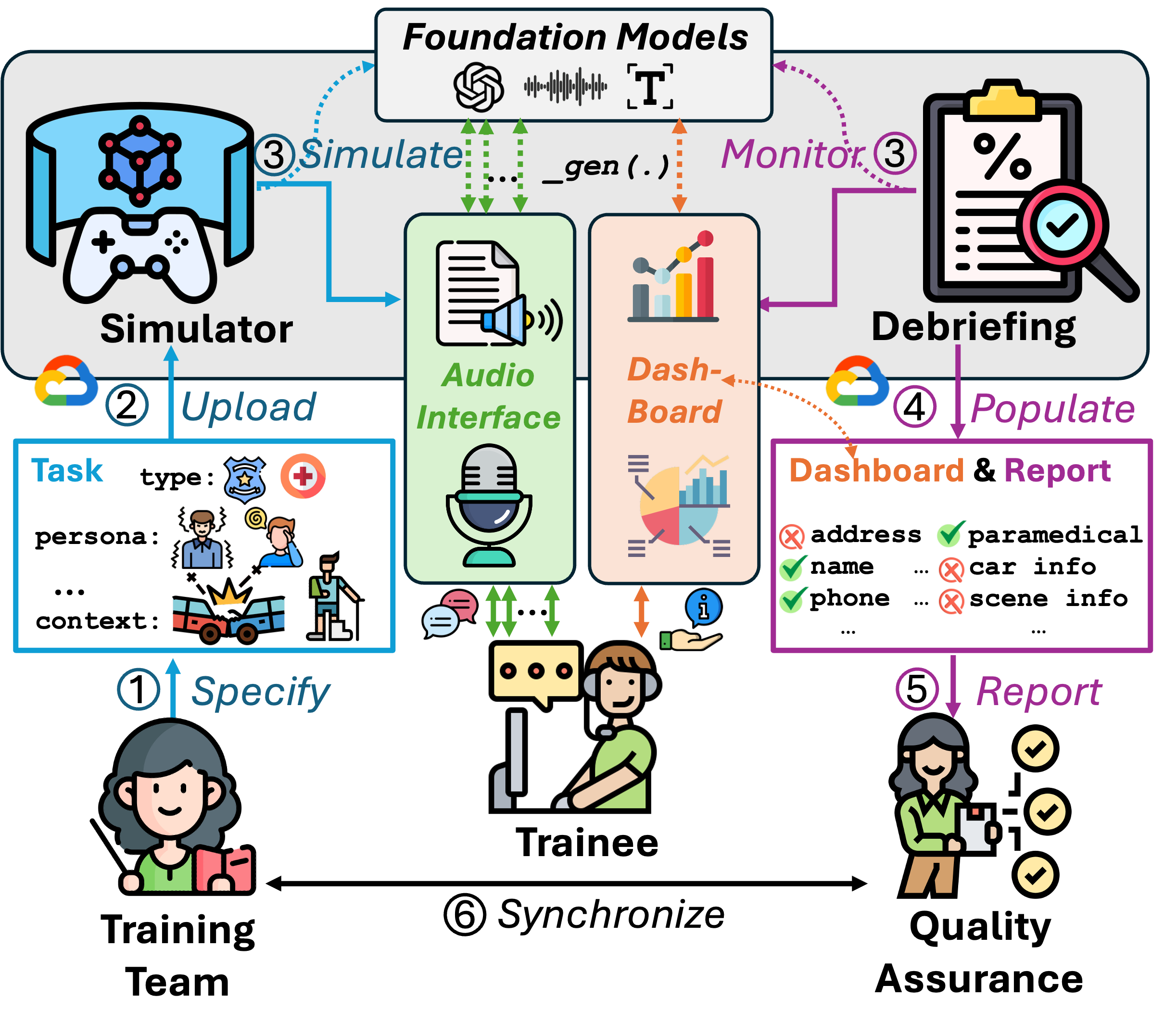}
  \caption{System operational workflow in six steps: (1) training team specifies tasks, (2) uploads to cloud platform, (3) foundation models role play caller and interact with trainees (via audio interfaces) while monitoring trainee performance (via dashboard), (4) automated debriefing populates performance reports, (5) quality assurance reviews outcomes, (6) results synchronization with training team for curriculum planning.}
  \label{fig:overview}
\end{figure}

The system automates two core training functions with GenAI: caller simulation~\cite{chen2025sim911} and performance assessment~\cite{chen2025logidebrief}. Training personnel configure scenarios by selecting from 57 incident types (e.g., cardiac arrest, structure fire, domestic disturbance), scenario contexts (e.g., severe weather, public gathering), and 100 caller persona combinations incorporating age, emotional state, language proficiency, and vulnerability factors such as cognitive impairment or mental health crisis. These parameters upload to cloud services~\cite{gcp} where foundation models, such as GPT-5.2~\cite{openai_gpt4o_audio_preview_2024}, generate realistic caller behavior including appropriate emotional dynamics, background environmental sounds, and physiologically plausible distress patterns.
Trainees engage through standard telephony interfaces identical to operational equipment, experiencing voice conversations similarly from live calls. During sessions, the system continuously monitors protocol compliance against agency standards through GenAI verification. These requirements encompass temporal constraints (e.g., obtaining address within first three turns), mandatory information gathering sequences, conditional protocols triggered by specific scene characteristics, and communication quality standards. Upon completion, automated debriefing generates performance reports with explanatory rationale for each completed or missed protocol step, highlighting both successful actions and improvement areas with specific examples from the conversation.
Quality assurance personnel validate outcomes by cross-referencing system assessments against audio recordings and protocol documentation before synchronizing results to training staff for curriculum adjustments and individual coaching. Figure~\ref{fig:overview} illustrates this end-to-end workflow.




\subsection{Design, Develop, and Deploy with Human in the Loop}
\label{sec:ddd}

\begin{figure}[h]
    \centering
    \includegraphics[width=\linewidth]{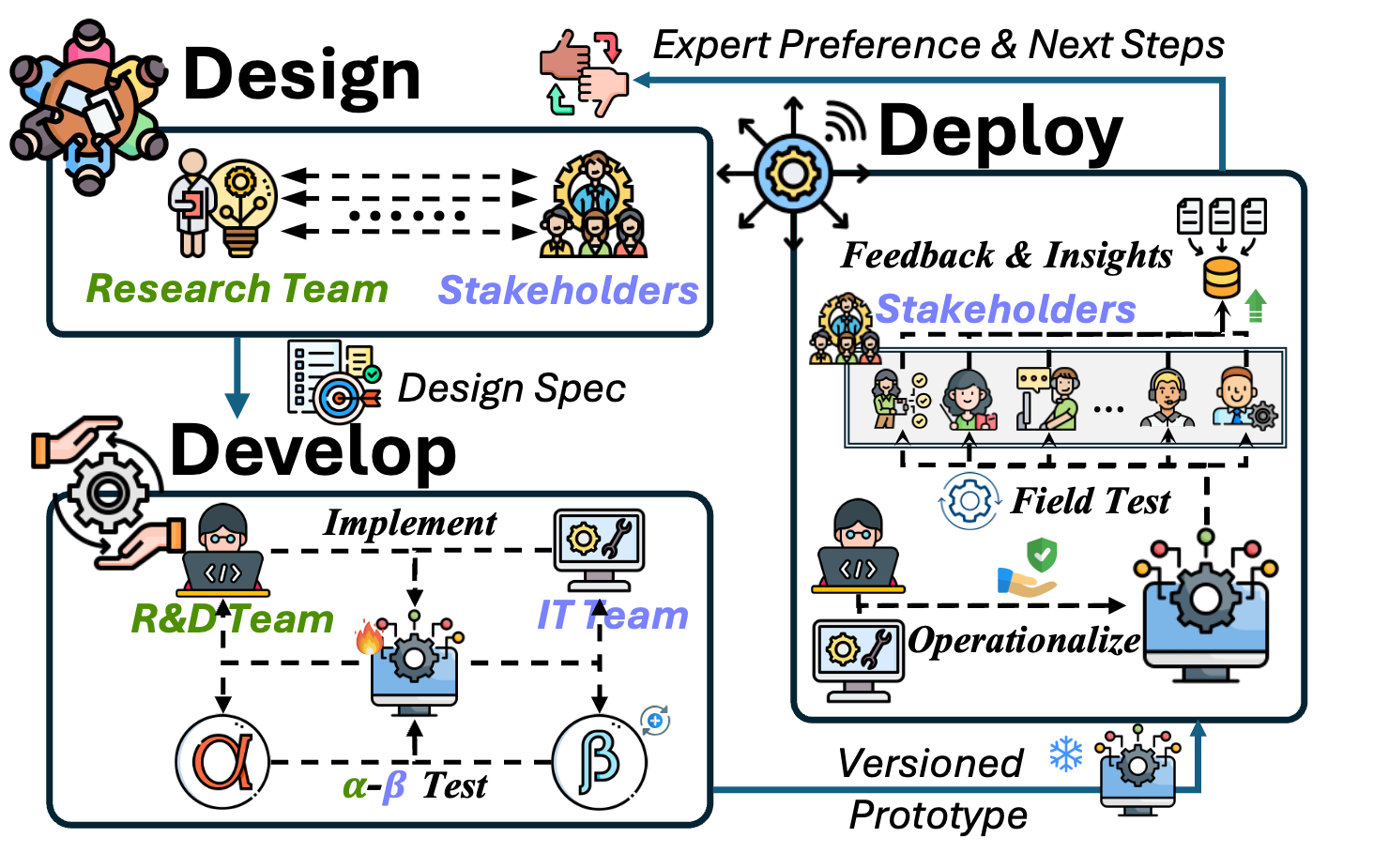}
    \caption{Continuously iterative design-develop-deploy workflow. Three phases operate concurrently with all around stakeholder participation throughout.}
    \label{fig:DDD}
\end{figure}

Our deployment diverges from sequential AI development by operating all following three phases concurrently rather than linearly (Figure \ref{fig:DDD}): (1) \textit{\textbf{Design}} proceeds through collaborative sessions where stakeholders validate system behaviors against institutional practice. Weekly routine meetings with diverse stakeholders (administrators, IT personnel, call-takers, training coordinators, quality assurance staff, and trainees) address system troubleshooting, feature prioritization, and operational integration challenges. On-site visits enable researchers to observe authentic training workflows, understanding how trainees progress through certification stages and how quality assurance personnel conduct manual call reviews. Listening sessions provide structured forums where government partners articulate logistical constraints around training schedules, workstation allocation, and staff availability. Workshops bring domain experts and researchers together to identify operational challenges and collaboratively evaluate technical solutions, ensuring proposed capabilities align with institutional feasibility. 
(2) \textit{\textbf{Development}} operates through dual-track implementation supported by both research and governmental IT teams. $\alpha$ deployment across controlled environments with 18 instrumented workstations enables rapid iteration based on immediate feedback from training coordinators and early adopters. $\beta$ rollout systematically expands to diverse user groups in staged increments, validating stability and usability before production integration. This phased approach allows technical refinement while building institutional confidence through demonstrated reliability at each scale. 
(3) \textit{\textbf{Deployment}} maintains continuous feedback channels where operational users submit reports through integrated interfaces at any time during or after training sessions. These reports trigger in-time triage by quality assurance personnel, with critical issues escalating to weekly development sprints for rapid patching. This responsiveness enables the system to adapt to emerging edge cases discovered only through sustained real-world use.


\subsection{Addressing Potential Ethical Concerns}
This study received approval from both institutional review board (IRB) and government agencies. All data collection adhered to strict ethical protocols: the system operates exclusively for training purposes with no access to live emergency calls, all participants provided informed consent, and personnel could opt out of AI-assisted training sessions at any time without penalty. Training and operational systems maintained strict separation, with all recorded data anonymized through hashed identifiers. The government agency co-owned all research data and participated in analysis design, ensuring institutional control over sensitive operational information. No personally identifiable information was collected, and audio recordings were retained only for quality assurance validation before secure deletion.

\section{Experiences \& Lessons Learned}
Scaling to operational use revealed four patterns critical for AI deployment in high-stakes government settings. We present lessons learned, emerged from longitudinal deployment and paired with actionable practices for future research.



\subsection{Bridging Knowledge Gaps Through Iterative Delivery}

\begin{quote}
    \textbf{Myth:} \textit{AI researchers can specify system requirements upfront by interviewing domain experts.}

    \textbf{Fact:} \textit{In deployments, neither AI teams understand operational constraints nor domain experts understand AI capabilities. Iterative delivery of minimal functioning systems enables mutual learning through concrete artifacts rather than abstract specifications.}
\end{quote}

\subsubsection{Observations} 

Deploying AI in specialized government domains faces a fundamental but common knowledge asymmetry: \textit{the research team lacks operational context while agency personnel lack AI knowledge} \cite{ehsan2023charting}. In our first meetings, government partners described needing ``\textit{more realistic training scenarios}'' but struggled to define what ``realistic'' meant without experiencing AI-powered simulation. We as researchers focused on incident type coverage, assuming more categories meant better training. Only when domain experts interacted with early prototypes did the critical gap emerge: caller diversity mattered more than incident variety. A heart attack scenario with a calm, native English-speaking caller trains fundamentally different skills than the same emergency with a panicked, non-native English speaker. Over 70 percent of real calls involve such vulnerable populations~\cite{mittal2025underserved, sasson2012neighborhood}, yet traditional role-playing cannot consistently simulate these variations. Without working systems, neither researchers nor practitioners identified caller diversity as a primary design constraint.

\subsubsection{Approach and Outcomes} 

We delivered a minimal system within one month using multi-modal LLMs~\cite{openai_gpt4o_audio_preview_2024} for caller role-play~\cite{brown2020language, wu2024role, dasswain2024teacher}. This working prototype transformed abstract discussions into concrete refinement. Training coordinators no longer described ``realistic scenarios'' in vague terms; they brought actual call clips, pointing to specific moments: ``\textit{Can we capture this kind of distress? Speech pattern? Background chaos?}'' Over six months through 27 meetings, 3 site visits, 2 listening sessions, and 1 workshop, something remarkable happened. Domain experts shifted from skepticism about AI capabilities to proposing AI-enabled solutions they couldn't have imagined initially. One training coordinator, initially doubtful that AI could simulate genuine emotional distress, became a strong advocate after hearing the system role-play a panicked non-English speaker: ``\textit{I didn't think AI could sound this real. We (as role-players) can't maintain this consistency across dozens of (simulated) calls.}'' Coverage expanded from 57 incident types to over 200 scenarios and from basic simulation to 100 caller personas incorporating age, emotion, language, and vulnerability, not because we added categories, but because domain experts learned to articulate what ``realistic'' actually meant through hands-on experience. Meanwhile, researchers discovered operational realities invisible in documentation: trainees needed graduated exposure to caller complexity, certain vulnerability combinations were systematically underrepresented in traditional training but matters a lot in real-world call-taking services. \textit{Both sides reached a convergent solution neither could have specified initially. Domain experts evolved from requirement providers into co-architects who jointly prioritized development. Researchers evolved from feature builders into operational partners who understood why features mattered}.

\subsubsection{Actionable Practices} For AI in governmental and high-stakes contexts: (1) \textit{Prioritize early delivery of minimal viable functionality} over comprehensive specifications, establishing shared artifacts that enable concrete feedback rather than abstract speculation. (2) \textit{Structure collaboration to cultivate co-architects} rather than auditors by involving domain experts in validation cycles, design reviews, and priority decisions throughout development rather than only at requirements and acceptance phases. (3) \textit{Instrument and celebrate bidirectional learning} as a deployment gain, tracking how operational insights reshape technical approaches while AI literacy enables stakeholders to articulate refined requirements they could not initially envision.

\subsection{Enhancing Rigor with Formal Methods}

\begin{table*}[t]
\footnotesize
\centering
\renewcommand{\arraystretch}{1.15}
\setlength{\tabcolsep}{6pt}

\caption{Examples of runtime checks encoded with LLM-backed \texttt{DETECT} predicate. Last six examples are conditioned on the context. All $\tau$ are adaptable hyper-parameters for different call-taking requirements.}
\label{tab:detect_examples}

\vspace{5pt}
\begin{tabular}{|m{6cm}|m{12cm}|}
\hline
\textbf{Natural Language Rules} & \textbf{Formalized Specifications} \\ \hline \hline

Call-taker asks for the address in the first $\tau_1$ turns. &
$\texttt{DETECT}\big(\omega_a^{[0,\tau_1]}, \text{`ask address'}\big)$ \\ \hline

Caller provides full name or phone number. &
$\texttt{DETECT}\big(\omega_b^{[0,T]}, \text{`provide full name / phone number'}\big)$ \\ \hline

Call-taker follows up on provided name / phone. &
$\Box_{[0,T]}
\Big(
    \omega_b(t) \models \text{`provides name / phone'}
    \;\rightarrow\;
    \texttt{DETECT}\big(\omega_a^{[t,\,t+\tau_2]}, \text{`follows up on name / phone'}\big)
\Big)$ \\ \hline

Before call end, call-taker verifies address again. &
$\texttt{DETECT}\big(\omega_a^{[T-\tau_3,\,T]}, \text{`ask address'}\big)$ \\ \hline

Scene safety info is obtained. &
(\textit{if the scene is potentially unsafe$\rightarrow$}) $\texttt{DETECT}\big(\omega_{ab}^{[0,\tau_4]}, \text{`scene safety info obtained'}\big)$\\ \hline

Caller is warned not to use energized equipment. &
(\textit{if an odor is reported}$\rightarrow$) $\texttt{DETECT}\big(\omega_a^{[0,\tau_5]}, \text{`warn caller not to use energized equipment'}\big)$ \\ \hline

Call-taker provides CPR instruction steps. &
(\textit{if patient is not breathing normally}$\rightarrow$) $\texttt{DETECT}\big(\omega_a^{[0,\tau_6]}, \text{`instructs [\textit{CPR step1, step2, ...}]'}\big)$\\ \hline

Call-taker asks for license plate and color. &
(\textit{if a vehicle is involved}$\rightarrow$)  $\texttt{DETECT}\big(\omega_a^{[0,\tau_7]}, \text{`ask for vehicle description'}\big)$\\ \hline

Caller is warned not to move it manually. &
(\textit{if roadway hazard is blocking traffic}$\rightarrow$)
$\texttt{DETECT}\big(\omega_a^{[0,\tau_8]}, \text{`warn caller not to move hazard'}\big)$ \\ \hline

Call-taker collects age and gender for EMS record. &
(\textit{if patient is conscious}$\rightarrow$) 
$\texttt{DETECT}\big(\omega_a^{[0,\tau_9]}, \text{`ask for patient demographics'}\big)$\\ \hline

\end{tabular}
\end{table*}

\begin{quote}
    \textbf{Myth:} \textit{Foundation models such as LLMs can reliably handle complex, end-to-end reasoning in safety-critical settings with prompt engineering alone.}

    \textbf{Fact:} \textit{In safety-critical AI, LLMs face severe limitations with long, intertwined contexts. Hybrid architectures that pair formal verification with modularized LLM predicates provide rigor while still leveraging generative capabilities.}
\end{quote}

\subsubsection{Observations}
AI systems in governmental contexts must satisfy strict procedural requirements while processing complex, evolving interactions. Emergency call debriefing exemplifies this tension: \textit{assessments must verify compliance against extensive protocol documentation spanning multiple agencies, temporal constraints, and conditional scene-dependent logic}. Pure LLM approaches systematically failed in this setting, with preliminary experiments showing consistent performance degrading over long contexts, e.g., Llama 3.2 \cite{touvron2023llama} dropped from 78\% solve rate to 42\% while the context (crammed with call transcripts and debriefing rules) only increased from 1,800 to 3,000 tokens. Supplying full context either exceeded model capacity or produced intertwined evidence where reasoning became unstable~\cite{liu2024lost, vodrahalli2024michelangelo}. Chain-of-thought prompting (CoT)~\cite{wei2022chain} and retrieval-augmented generation (RAG)~\cite{lewis2020retrieval} offer only marginal gains and cannot reach the verification rigor~\cite{weng2024mastering,dong2024exploring} required for deployment, where errors risk reinforcing incorrect protocols with irreversible consequences~\cite{arcuschin2025chainofthought}.

\subsubsection{Approach and Outcomes}
We formulate 9-1-1 call debriefing as runtime verification over a signal trace $\omega_{ab}$, where $\omega_a$ and $\omega_b$ denote the ordered utterance sequences of the call-taker and caller, and $\omega_{ab} = \omega_a \cup \omega_b$ with temporal ordering preserved. In collaboration with domain experts, we translate 1,651 emergency protocol requirements into Signal Temporal Logic (STL) specifications. LLMs are restricted to narrow, modularized predicates. Formally, for an LLM-embedded boolean \texttt{DETECT} function\footnote{More predicates are defined and developed for real-world usage; here \texttt{DETECT} only serves as an exemplar clarification}:

\begin{equation}
\texttt{DETECT}(\omega, \text{action}) \coloneqq \Diamond_{[0,T]} \underbrace{ \big(\omega(t) \models \text{action} \big)}_{\text{LLM Prompts}}
\end{equation}
If checking ``if the call-taker asked for address in the first 3 turns'', it can be rewritten as $\texttt{DETECT}\big(\omega_a^{[0,\tau]}, \text{`ask address'}\big)$, where $\tau=3$. More examples can be found in Table \ref{tab:detect_examples}.

Our evaluation across 1,244 real-world calls demonstrated substantial rigor improvements. The formal methods approach achieved 94.3-95.9 percent agreement with expert assessments compared to 87.8 percent for the strongest pure LLM baseline (In-context learning with CoT \cite{wei2022chain} and RAG \cite{lewis2020retrieval}). Critically, the hybrid architecture provides verifiable correctness guarantees: \textit{every assessment traces to specific signals, temporal constraints, and logical rules that domain experts validated during collaborative specification development}. This transparency builds institutional trust essential for operational adoption in governmental contexts where accountability and auditability are non-negotiable.


\subsubsection{Actionable Practice} 
For AI in governmental and high-stakes contexts requiring procedural compliance: (1) \textit{Identify domain constraints amenable to formal specification}, particularly temporal requirements, conditional logic, and sequential dependencies that resist reliable extraction through pure LLM reasoning. (2) \textit{Collaborate with domain experts to encode requirements as human-readable formal specifications} independently verifiable from implementation, establishing ground truth that builds institutional trust through transparency. (3) \textit{Restrict LLM responsibility} to well-defined predicates with narrow scope and verifiable inputs/outputs, avoiding open-ended reasoning over complex integrated contexts. (4) \textit{Design architectures where failures are traceable} to specific components, enabling systematic refinement rather than opaque model retraining when errors occur.

\subsection{Building Resilience via Triangulated Feedback}
\label{sec:resilience}
\begin{quote}
    \textbf{Myth:} \textit{User-reported errors reliably indicate technical failures requiring system fixes.}

    \textbf{Fact:} \textit{In high-pressure training environments, a substantial proportion of error reports reflect psychological stress responses rather than system faults. Triangulated validation distinguishes genuine failures from misattributed performance.}
\end{quote}

\subsubsection{Observations}

\begin{figure}
    \centering
    \includegraphics[width=\linewidth]{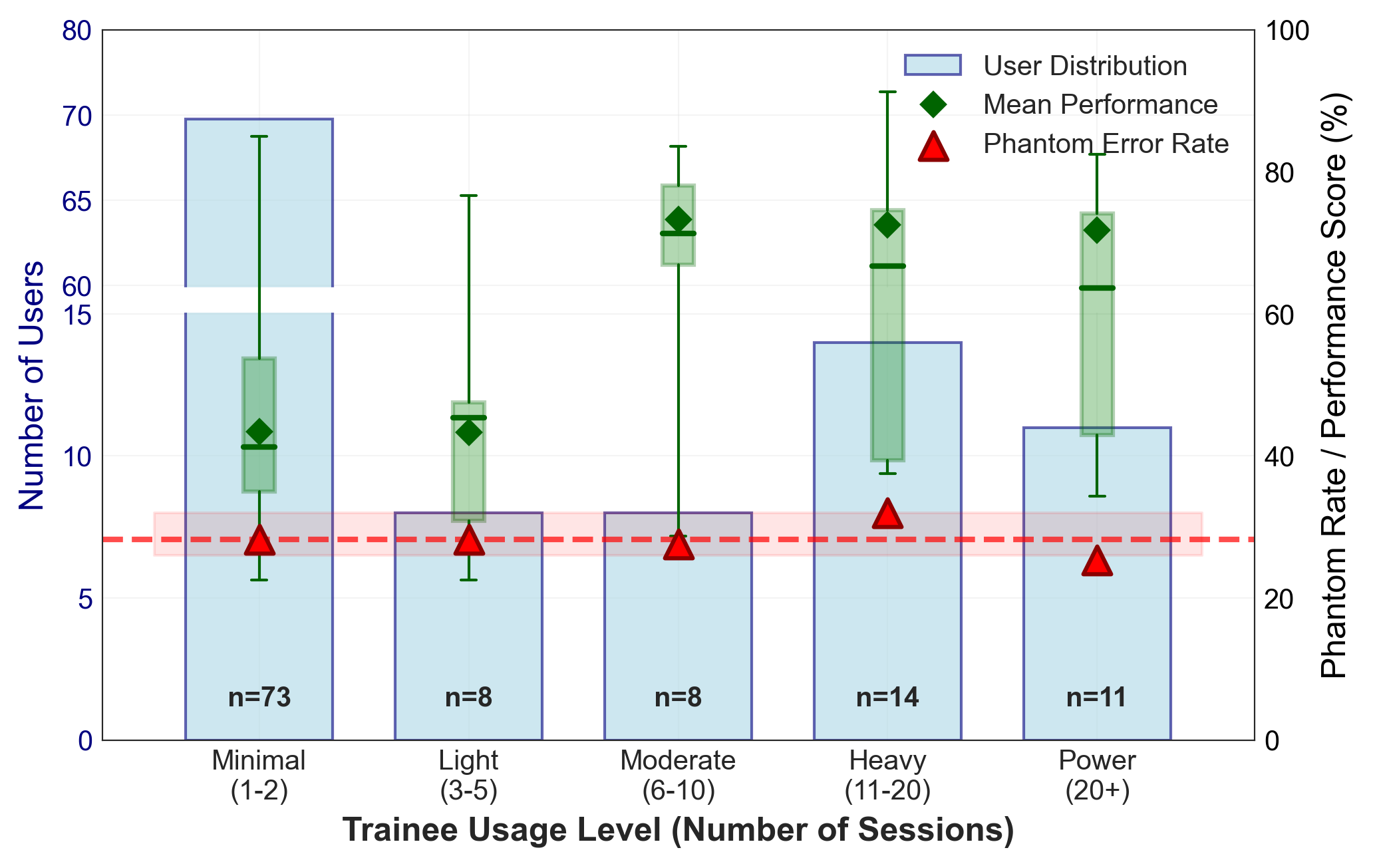}
    \caption{Dispute rates (noted as `\textit{phantom error}') of user attributed mistakes to system different across experience and performance levels (114 trainees, 1,120 completed sessions).}
    \label{fig:phantom}
\end{figure}

When training systems perform evaluative functions under pressure, \textit{users experiencing difficulty may attribute failures to technology rather than their own performance as a psychological defense mechanism}~\cite{kubota2014stress}. This creates a critical dilemma: responsive organizations must investigate user concerns to maintain trust, yet treating every report as technical failure risks destabilizing working systems through unnecessary modifications. Traditional binary approaches either dismiss user feedback as unreliable or accept all reports uncritically, both undermining long-term adoption.
One trainee reported a ``system error'' claiming they had verified the caller's address but received no credit. Session recordings showed the trainee asked for an address and heard ``123 Main Street,'' but immediately moved on without required verification: confirming spelling, checking cross-streets, or validating against geographic information. The trainee genuinely believed asking equaled verifying under stress. This illustrates how 28.24\% of error reports are misattributed to system failures when evidence confirms trainee mistakes (Figure~\ref{fig:phantom}). Manual review of recordings, activity logs, and debriefing outputs reveals this pattern persists across experience levels, indicating psychological stress responses rather than misunderstandings training could resolve~\cite{wischnewski2023measuring, buccinca2021trust, ma2023who}.

\subsubsection{Approach and Outcomes}
We establish triangulated feedback loops preventing premature system modifications. Rather than direct user-to-developer channels, we loop in quality assurance personnel as independent validators. When trainees report errors, the system surfaces correlated evidence including timestamped recordings, activity logs, and assessment rationale. Quality assurance staff independently validate claims against operational standards before issues enter development prioritization. This structure serves dual purposes: \textit{distinguishing genuine technical failures from stress-induced mis-attributions}, and identifying when reported ``errors'' actually reveal system behaviors correctly aligned with training objectives that users find uncomfortable. This triangulated feedback loop enables confident and resilient system evolution. And it prevents wrong-direction updates that would have degraded system effectiveness. In our scenario, 28.24\% of reported ``errors'' reflect the system accurately enforcing protocols that trainees find difficult, precisely the productive struggle necessary for skill development. Without validation, developer response to eliminate user frustration would have weakened training rigor.

\subsubsection{Actionable Practices}
For AI in governmental and high-stakes training contexts: (1) \textit{Build evidence-linked review interfaces} that surface correlated operational artifacts (logs, recordings, state traces) enabling independent validation by domain experts without requiring technical expertise. (2) \textit{Establish triangulated feedback} that routes user reports through domain expert validation before development prioritization. Explicitly categorize each resolution as either genuine technical failure or stress-induced misattribution, enabling pattern tracking over time.


\subsection{Designing Explanations for Constructive Learning}

\begin{quote}
    \textbf{Myth:} \textit{Effective training systems should minimize user frustration and maximize immediate satisfaction.}

    \textbf{Fact:} \textit{High-stakes training requires calibrated difficulty inducing productive struggle. Optimal learning emerges from scenarios that challenge users while providing constructive feedback acknowledging strengths alongside improvement areas.}
\end{quote}

\subsubsection{Observations}

\begin{figure}
    \centering
    \includegraphics[width=\linewidth]{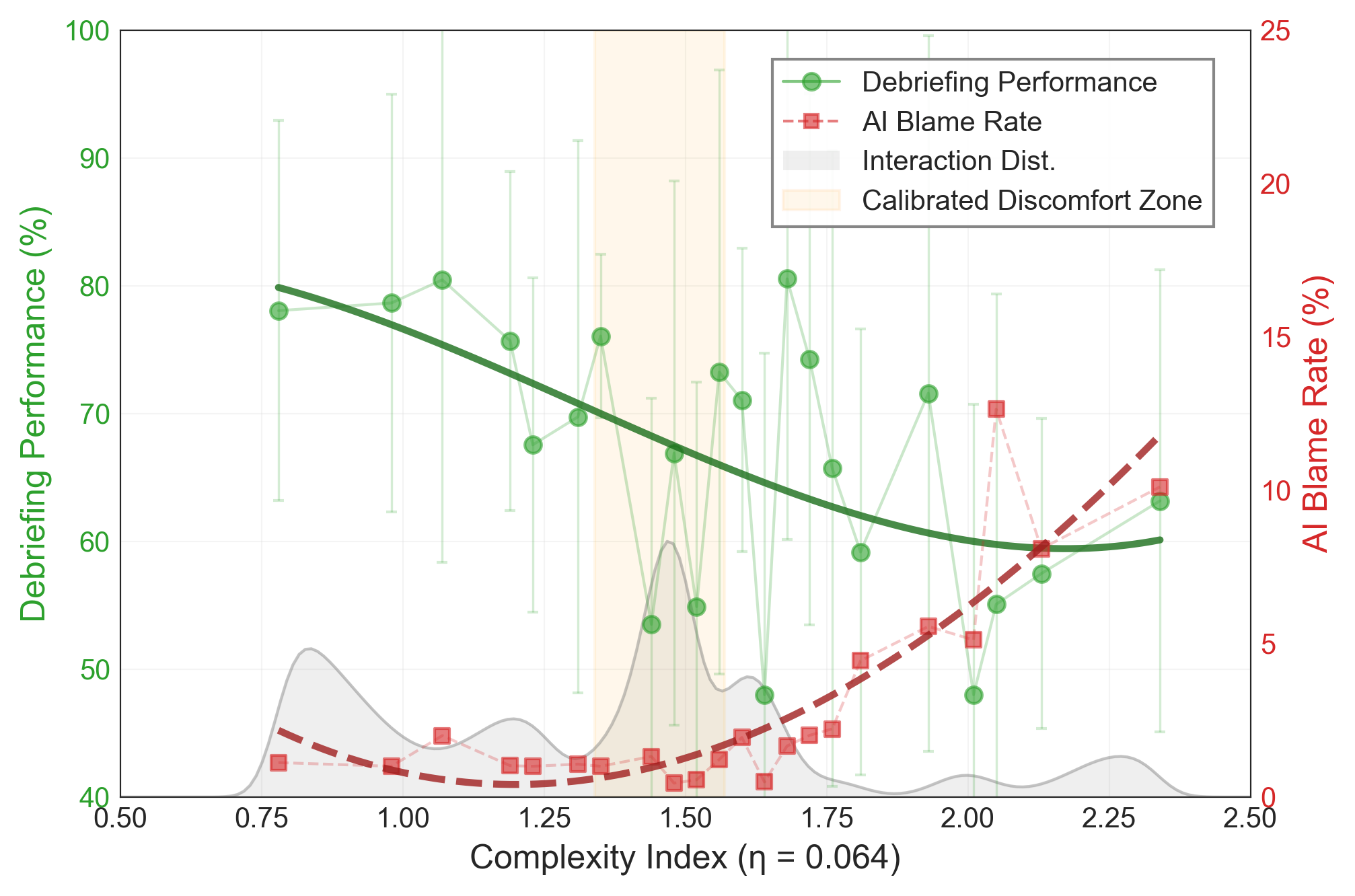}
    \caption{Task complexity versus performance and dispute rates (940 sessions with 12+ turns). Green lines indicate the debriefing performance, i.e., how well the trainees did during simulation. Red dashed lines indicate the AI blame rate, i.e., when trainees mis-attribute their errors to AI.}
    \label{fig:discomfort}
\end{figure}

AI systems in high-stake training contexts face a widely-recognized tension: \textit{training must be challenging to prepare users for real-world demands, yet excessive stress undermines skill acquisition} \cite{kapur2008productive}. As one training coordinator explained: ``\textit{The most valuable aspect of difficult scenarios is to build resilience before the chaos of a real emergency. In those critical moments, real lives depend on it.}''
We define Complexity Index as $\text{CI} := \eta \times N + D + C$, where $\eta = 1/15.6 =0.064$, represents the inverse average number of required actions across all protocols under our operational context. $N$, $D$, $C$ are the numbers of protocol requirements (call-taker's todo), departmental coordination (police, fire, medical), and activated caller profiles (e.g., emotion, age, etc).
Analysis of 940 longer training sessions (Figure~\ref{fig:discomfort}) reveals systematic relationships between scenario complexity and performance. Moderate negative correlation exists between complexity and scores (r = -0.66, p = 0.001), with performance decreasing 6.7 percentage points per complexity unit. Dispute rates (also referenced in Section~\ref{sec:resilience}) where users attribute failures to system errors show stronger positive correlation with complexity (r = 0.75, p $<$ 0.001), increasing 2.6 percentage points per unit. Domain experts confirmed that stress levels at moderate complexity mirror real emergency conditions, validating deliberate challenge cultivation as essential preparation.

\subsubsection{Approach and Outcomes}
Early system feedback followed common patterns: summarizing interactions and enumerating protocol deficiencies. This deficit-focused approach amplified stress without providing learning scaffolds. Users encountering complex scenarios with multiple missed requirements received extensive lists of failures without acknowledgment of successfully executed steps. Stakeholder feedback revealed this approach increased frustration and reduced motivation precisely when productive struggle is most valuable for skill development. We restructure feedback to incorporate explicit strengths acknowledgment. For each protocol deficiency flagged, the system highlights successfully completed requirements and effective communication techniques demonstrated during the same session. This balanced approach maintains assessment accuracy while reframing struggle as productive learning rather than comprehensive failure. This constructive feedback architecture maintains necessary difficulty while providing psychological scaffolding through the learning process. Post-revision analysis showed session completion rates (users can opt out of AI simulated training any time) increased from 76 percent to 89 percent among users encountering high-complexity scenarios, with qualitative feedback indicating reduced frustration despite unchanged scenario difficulty. The approach validated that AI training systems can maintain rigorous standards while supporting users through `\textit{productive struggle}' \cite{marraffino2021adapting, marraffino2021adapting}. A calibrated discomfort zone in our case emerged at moderate complexity levels (Complexity Index $\in$ [1.34, 1.57], conceptually 1.34-1.57 times \textit{`harder'} than average) where performance remains achievable yet challenging, with constructive feedback enabling sustained engagement in this optimal learning range.


\subsubsection{Actionable Practices}
For AI in governmental and high-stakes contexts requiring demanding skill development: (1) \textit{Design for calibrated difficulty} rather than user comfort, formalizing complexity estimation through quantifiable dimensions that enable controlled progression within productive struggle zones. (2) \textit{Implement balanced feedback architectures} that explicitly acknowledge successful actions alongside areas for improvement, maintaining motivation through challenging scenarios by framing struggle as productive learning rather than failure.

\section{Related Work}

Our work builds on research spanning responsible AI frameworks, government AI adoption, and socio-technical systems design.
Studies of \textbf{AI adoption in public sectors} identify infrastructure limitations, organizational resistance, and procurement constraints as persistent barriers, with successful deployments requiring sustained stakeholder engagement across organizational hierarchies and executive-level support~\cite{mergel2024implementing, chen2023cityspec, chen2022intelligent, chen2022cityspec}. These studies further note that pilot-stage success rarely transfers directly to operational settings without governance alignment, workflow integration, and long-term maintenance planning~\cite{mergel2024implementing}.
Research on \textbf{trust in high-stakes AI systems} demonstrates that psychological factors such as stress responses, attribution biases, and perceived accountability drive reliance patterns independent of technical performance, with both overtrust and undertrust leading to system abandonment~\cite{lee2004trust, alufaisan2024systematic}. Subsequent work shows that calibrated trust depends on explanation quality, feedback loops, and observable system behavior under edge cases rather than aggregate accuracy metrics alone~\cite{lee2004trust, alufaisan2024systematic}. While \textbf{responsible AI frameworks} provide governance principles emphasizing transparency, fairness, accountability, and safety~\cite{jobin2019global}, they lack operational guidance for sustained deployment under embedded constraints such as how to operationalize transparency when users lack technical literacy, or how to maintain accountability when systems span multiple organizational jurisdictions~\cite{saxena2024democratizing, delgado2023participatory}.
Work on \textbf{AI-powered training systems} in emergency response validates effectiveness for skill acquisition through simulation-based learning and automated assessment~\cite{beham2023artificial, hong2025exploring}, yet focuses on training outcomes rather than deployment challenges, lifecycle maintenance, and cross-agency interoperability requirements~\cite{beham2023artificial, hong2025exploring}. \textbf{Socio-technical systems theory} emphasizes joint optimization of technical and social elements through participatory design and iterative refinement~\cite{baxter2011socio, pasmore2019reflections}, with empirical studies showing organizational factors often dominate technical performance in determining adoption success~\cite{mumford2006story}. Recent socio-technical analyses of AI systems also highlight the need for continuous monitoring, governance adaptation, and role redefinition after deployment rather than one-time integration efforts~\cite{pasmore2019reflections}.
However, existing frameworks remain conceptual without concrete implementation guidance or reproducible deployment patterns under real operational constraints.
Our deployment extends this foundation by translating abstract principles into replicable practices grounded in empirical evidence from production deployment, including workflow embedding, stakeholder-aligned evaluation, and iterative post-deployment refinement.

\section{Limitation and Discussion}
This study reports a sustained deployment within a single municipal emergency communications agency. Although this setting provides ecological validity under real operational constraints, the findings remain shaped by local protocols, staffing structures, infrastructure, and training culture. Future work should examine deployments across agencies and regions to identify which design practices generalize beyond MNDEC.

The current scope also focuses on call-taking training, including simulated caller interaction, protocol-oriented assessment, and debriefing. It does not yet fully capture downstream dispatch coordination, multi-agency incident evolution, or transfer from simulated practice to live operational performance. Future evaluations should connect AI-supported training with longitudinal outcomes such as certification progress, retention, confidence calibration, live-call quality assurance, and supervisor assessment.

More broadly, this work raises a larger question for AI-augmented labor: how can AI-supported workflows preserve and enhance workers' agency, meaning, and purpose? Our deployment suggests that AI is most valuable when it scales expert knowledge without displacing human judgment. Trainers, quality assurance personnel, and domain experts remained central to defining realism, validating outcomes, and interpreting feedback. This points toward apprenticeship and reskilling models where AI supports adaptive practice, targeted feedback, and reflective learning rather than simple task automation. In public-safety work, the goal is not only faster procedural training, but the cultivation of human-essential skills: judgment under uncertainty, communication under stress, ethical reasoning, procedural discipline, and appropriate reliance on AI systems.

\section{Conclusion}



This paper reports lessons learned from deploying a GenAI-powered 9-1-1 call-taking training system in a municipal emergency communications center under real operational constraints.
Beyond model performance, our experience shows that sustained adoption in safety-critical public-sector settings is primarily shaped by delivery practices, governance structure, and human factors.
From longitudinal deployment, we identify four key takeaways:
(1) \textbf{iterative delivery with minimal working artifacts} is essential for resolving specification ambiguity and enabling domain experts to meaningfully participate as co-designers rather than post hoc auditors.
(2) \textbf{scoping LLM responsibility through formalized constraints} improves auditability and reliability, allowing procedural requirements to be verified independently of generative components.
(3) \textbf{evidence-linked and expert-validated feedback loops} are necessary to prevent stress-induced misattribution from driving incorrect system updates, preserving training rigor over time.
(4) \textbf{calibrated challenge and constructive feedback} support effective skill development by maintaining necessary difficulty without discouraging trainees.
Together, these findings suggest that deploying GenAI in safety-critical government contexts requires treating socio-technical integration, validation pathways, and post-deployment iteration as first-class system components.
This work provides actionable guidance for researchers and practitioners seeking to move responsible AI from principles to practice in real-world public-sector systems.

\section*{Acknowledgments}

This work was supported in part by the U.S. National Science Foundation under Grants 2427711 and 2443803, the Google Academic Research Award, and the U.S. Department of Education under Grant R305C240010.    
The opinions, findings, conclusions, or recommendations expressed in this material are those of the author(s) and do not necessarily reflect the views of the sponsoring agencies. 
The authors thank Yilin Liu for his contributions to the final preparation of this manuscript and for presenting this work.



\bibliographystyle{plain}
\bibliography{refs}

\end{document}